\pdfoutput=1
\documentclass[a4paper,11pt]{article}

\usepackage[T1]{fontenc}
\usepackage{lmodern}
\usepackage[margin=2.5cm]{geometry}

\usepackage{amsmath, amssymb, amsthm, mathtools}
\usepackage{thm-restate}
\usepackage{aligned-overset}
\usepackage{enumitem}
\usepackage{xspace}
\usepackage[ruled,vlined]{algorithm2e}
\usepackage{tikz}
\usepackage{pgfplots}
\pgfplotsset{compat=1.18}
\usepackage[hidelinks]{hyperref}
\usepackage[capitalise, noabbrev]{cleveref}
\crefname{algocf}{Algorithm}{Algorithms}
\Crefname{algocf}{Algorithm}{Algorithms}

\definecolor{blue}{HTML}{0072B2}
\definecolor{lightblue}{HTML}{56B4E9}
\definecolor{rose}{HTML}{CC79A7}
\definecolor{orange}{HTML}{D55E00}
\definecolor{yellow}{HTML}{F0E442}
\definecolor{green}{HTML}{009E73}

\theoremstyle{plain}
\newtheorem{theorem}{Theorem}
\newtheorem{lemma}[theorem]{Lemma}

\newtheorem{definition}[theorem]{Definition}
\newtheorem{observation}[theorem]{Observation}
\crefname{observation}{Observation}{Observations}
\Crefname{observation}{Observation}{Observations}

\SetKwIF{If}{ElseIf}{Else}{if}{\!:}{else if}{else}{endif}
\DontPrintSemicolon

\newcommand{\R}{\mathbb{R}}
\newcommand{\pos}{\mathrm{pos}}
\DeclareMathOperator{\sgn}{sgn}
\DeclareMathOperator{\serve}{serve}
\newcommand{\servingTime}[1]{\ensuremath{\serve(r_{#1})}}
\newcommand{\LB}{\mathrm{LB}}
\newcommand{\x}{\boldsymbol{x}}
\newcommand{\y}{\boldsymbol{y}}
\renewcommand{\epsilon}{\varepsilon}

\newcommand{\ALG}{\text{\normalfont\scshape Alg}\xspace}
\newcommand{\OPT}{\text{\normalfont\scshape Opt}\xspace}
\newcommand{\leftfirst}{\text{\normalfont\scshape LeftFirst}\xspace}
\newcommand{\ouralg}{\text{\normalfont\scshape LeftGuard}\xspace}
\newcommand{\movel}{\text{\normalfont\scshape MoveLeft}\xspace}
\newcommand{\mover}{\text{\normalfont\scshape MoveRight}\xspace}

\title{Improved Bounds for Online TSP on the Half-Line\thanks{An extended abstract of this paper appeared in the proceedings of the 20th Scandinavian Symposium on Algorithm Theory (SWAT 2026)~\cite{Baligacs2026}.}}
\author{
	J\'ulia Balig\'acs\thanks{University of Oxford, \texttt{jbaligacs@gmail.com}, \href{https://orcid.org/0000-0003-2654-149X}{ORCID 0000-0003-2654-149X}. Funded by the European Union (ERC, CCOO, 101165139). Views and opinions expressed are however those of the authors only and do not necessarily reflect those of the European Union or the European Research Council. Neither the European Union nor the granting authority can be held responsible for them.}
	\and
	Yann Disser\thanks{TU Darmstadt, \texttt{disser@mathematik.tu-darmstadt.de}, \href{https://orcid.org/0000-0002-2085-0454}{ORCID 0000-0002-2085-0454}.}
	\and
	Linda Thelen\thanks{TU Darmstadt, \texttt{thelen@mathematik.tu-darmstadt.de}, \href{https://orcid.org/0009-0006-1625-0567}{ORCID 0009-0006-1625-0567}.}
}

\date{}

\begin{document}

\maketitle

\begin{abstract}
	In the open online traveling salesperson problem, requests appear over time at different positions of a metric space.
	A single agent traveling at unit speed must serve all requests, by visiting their positions, with the objective of minimizing the completion time.
	We propose online algorithms for this problem for the case that the metric space is the half-line.
	First, we present a 2-competitive algorithm in which the server always either moves at unit speed or remains stationary, which improves on the previously best-known ratio of~2.04.
	We further observe that algorithms must sometimes move slower than unit speed to achieve competitive ratios below~2.
	We introduce a second algorithm that beats the barrier of~2 by carefully modulating its speed, and prove a tight bound of~1.75 on its competitive ratio.
	Finally, we slightly strengthen a known lower bound on the best-possible competitive ratio on the half-line from~$1.627$ to~$1.646$.

\end{abstract}

\section{Introduction}

We study the online traveling salesperson problem (TSP) on the half-line.
In this problem, requests of the form $r=(t,p)$ are revealed over time, where $t\in \R_{\ge 0}$ denotes the release time and $p\in \R_{\ge 0}$ the position of the request.
A single server, initially located at the origin and moving at speed at most~$1$, must serve every request by visiting position $p$ at some time not earlier than time~$t$.
In the online setting, the algorithm only learns about a request at its release time, whereas an off\-line algorithm knows the entire request sequence from time~$0$.
The objective is to minimize the makespan, i.e., the time when all requests are served.
Note that the offline setting does not coincide with the classical TSP, as the off\-line algorithm is still restricted to serving a request after its release time.
Furthermore, we consider the \emph{open} version of the problem, where, in contrast to the \emph{closed} version, the server \emph{does not} need to return to the origin.

Online TSP models dynamic routing scenarios arising, for instance, in robotic maintenance, cleaning, or collection tasks where jobs appear over time.
In many such settings, the depot lies at the boundary of a one-dimensional operating region, such as inspection along a track, pipeline, or corridor, so that the underlying metric space is the half-line, making it a natural case to study.
While the online TSP has been extensively studied in many variants~\cite{AsymmetricOnlineTSP,AusielloDemangeLauraPaschos/04,Ausiello/01,BlomKrumkePaepeStougie/01,JailletLu/11,JailletLu/14,JawgalMuralidharaSrinivasan/19,Krumke/02}, in particular for the underlying metric space being the real line~$\R$ or higher-dimensional Euclidean spaces~$\R^n$, we focus here on algorithms tailored specifically to the half-line~$\R_{\geq 0}$.
Although this setting appears simple at first glance, obtaining optimal strategies turns out to be surprisingly challenging.
The half-line has fundamentally different structural properties due to the lack of symmetry around the origin, and it seems that optimal strategies cannot be obtained by simple adaptations of known algorithms.
For instance, we will see that, unlike any algorithm previously proposed for online TSP, an optimal online strategy on the half-line must already start moving before the first request is revealed.

As usual in online optimization, we measure the quality of a (deterministic) algorithm in terms of \emph{competitive analysis}.
For an algorithm $\ALG$ and a request sequence $\sigma$, let $\ALG(\sigma)$ denote the completion time of $\ALG$ on $\sigma$, and let $\OPT(\sigma)$ denote the completion time of an optimal off\-line solution.
We say that $\ALG$ is \emph{$\rho$-competitive} if $\ALG(\sigma)\le \rho\,\OPT(\sigma)$ for every request sequence $\sigma$.
The \emph{competitive ratio of $\ALG$} is the infimum over all such $\rho$, and the \emph{competitive ratio of the problem} is the best ratio achievable by any online algorithm.

Throughout the paper, we assume without loss of generality that every request $r=(t,p)$ satisfies $t \ge p$.
Indeed, since the server moves at speed at most~1, even the off\-line optimum cannot reach position $p$ before time $p$, and is therefore unaffected by this assumption.
Moreover, an online algorithm could only benefit from learning about a request earlier.
Hence, this assumption is justified when the request sequence is constructed adversarially and does not affect the competitive ratio of the problem.

Finally, we introduce the following notational conventions.
Given an online algorithm~\ALG, we write $\pos(t)\in \R_{\ge 0}$ for the position of the server at time~$t$. For a request sequence~$\sigma$, we sometimes abbreviate the completion times $\ALG(\sigma)$ and $\OPT(\sigma)$ by \ALG and \OPT when the sequence is clear from context.
At any point in time, the server can move in (at most) two directions. We refer to the direction towards the origin as \emph{left} and to the direction away from the origin as \emph{right}.

\paragraph*{Our results}

We begin by introducing the algorithm \leftfirst (cf.~\Cref{alg:leftfirst}), which follows a very simple strategy: it prioritizes requests that lie to the server's left over those to its right. We show that this algorithm achieves a competitive ratio of~2, improving on the previously best-known ratio of~2.04~\cite{BjeldeDisserHackfeldEtal/20}.
We further observe that no algorithm restricted to moving either at full speed or not at all can achieve a better competitive ratio.

\begin{theorem}
	\label{thm:2-competitive}
	The algorithm \leftfirst is $2$-competitive for online TSP on the half-line.
	Moreover, this ratio is optimal among all algorithms in which the server is restricted to speeds~$0$ and~$1$.
\end{theorem}

In particular, the theorem shows that fundamentally new ideas are required to beat the barrier of~2.
Our main contribution is to propose such an idea in the form of a parameterized algorithm $\ouralg(\alpha)$ for $\alpha\in(0,1]$ (cf.~\Cref{alg:ouralgorithm}).
Its strategy can be summarized as follows.
The algorithm maintains at all times that $\pos(t)\le \alpha t$, enabling the server to reach requests to its left in a timely manner.
It assigns strict deadlines to such requests and prioritizes requests to its right if and only if
these deadlines can still be met.
Our main result is a careful analysis of this algorithm.
We show that, for $\alpha=0.75$, it achieves a competitive ratio of~1.75.
After identifying the most difficult instances for $\ouralg$, a key ingredient of our proof is an LP-duality argument that yields sharp performance bounds. Moreover, we prove that this analysis is tight in the sense that $\ouralg(\alpha)$ has competitive ratio at least~1.75 for every choice of~$\alpha$.

\begin{theorem}
	\label{thm:mainthm}
	$\ouralg(0.75)$ is 1.75-competitive for open online TSP on the half-line.
	Moreover, the choice $\alpha=0.75$ is optimal.
\end{theorem}

We complement our result with a general lower bound on the best-possible competitive ratio that slightly improves the known lower bound of 1.627~\cite{Lipmann/03}.

\begin{restatable}{theorem}{thmLowerBound}
	\label{thm:lower_bound}
	Every algorithm for open online TSP on the half-line has competitive ratio at least~1.6463.
\end{restatable}

\paragraph*{Related work}
While this is, to the best of our knowledge, the first work to specifically focus on open online TSP on the half-line, the problem has been studied before on other metric spaces.
Most notably, Bjelde et al.~tightly analyzed the problem on the line with a competitive ratio of~2.04~\cite{BjeldeDisserHackfeldEtal/20}.
Their upper bound also applies to the special case of the half-line and, prior to our work, was the best-known upper bound in this setting.
Their lower bound extends to general metric spaces and remains the best-known lower bound there.
The best-known algorithm for general metric spaces, introduced by Bonifaci and Stougie, achieves a competitive ratio of~$1+\sqrt{2}\approx 2.41$~\cite{BonifaciStougie/08}.

By contrast, the closed version of the problem has been tightly analyzed in all three cases: It has a competitive ratio of 2 on general metric spaces~\cite{Ausiello/01}, of 1.64 on the line~\cite{Ausiello/01,BjeldeDisserHackfeldEtal/20} and of~1.5 on the half-line~\cite{BlomKrumkePaepeStougie/01}.
The optimal algorithm for the closed version on the half-line follows a strategy that can be viewed as the ``opposite'' of our algorithm \leftfirst: That algorithm moves right whenever a request to the server's right is available and otherwise moves left.

Online TSP has also been investigated in several related settings, for instance when the metric space is a circle~\cite{JawgalMuralidharaSrinivasan/19}, when distances are asymmetric~\cite{AsymmetricOnlineTSP}, or when the request sequence satisfies additional structural assumptions~\cite{BlomKrumkePaepeStougie/01,Krumke/02}.

The online TSP can be viewed as a special case of the extensively studied \emph{online dial-a-ride problem}~\cite{AscheuerKrumkeRambau/00,wads,BirxDisser/20,BirxDisser/22,FeuersteinStougie/01,KrumkePaepePoensgenEtal/06,Lipmann/03}, where each request is a transportation request consisting of a release time, a starting position, and a destination.
In other words, the online TSP corresponds to the special case of online dial-a-ride in which starting position and destination always coincide.
Classical online dial-a-ride strategies include \textsc{Ignore}, which repeatedly recomputes a schedule for the currently unserved requests~\cite{Birx/20,Krumke0};
\textsc{Replan}, which greedily recomputes an optimal tour whenever a new request arrives~\cite{Ausiello/01,Birx/20,Krumke0};
and \textsc{Lazy}, which deliberately delays service to aggregate requests before executing a tour~\cite{wads,Lipmann/03}.

Other online variants of the TSP focus on different types of uncertainty.
In \emph{online graph exploration}~\cite{BaligacsDisserHeinrichSchweitzerExploration,Cosson24,DisserHackfeldKlimm/19,pruhs94}, the metric space is initially unknown and all vertices must be visited.
In \emph{universal TSP}~\cite{Christodoulou2017,Jia2005}, the goal is to find a fixed master tour that is competitive for any subset of active requests.
Recently, Abrahamsen et al.~\cite{Abrahamsen2024} introduced a model where requests are revealed one by one and must be irrevocably sorted to minimize total travel distance, rather than minimizing the makespan for requests arriving over time.

\section{Warm-up: moving at speed 1 or 0}

Before considering the problem in full generality, we first consider the easier case where the server is only allowed to move at full speed or not move at all.
We first observe a simple lower bound on the competitive ratio of this variant of the problem.

\begin{lemma}
	\label{lem:2-lower}
	If the server is only allowed to move at speed 1 or 0, the competitive ratio of open online TSP on the half-line is at least 2.
	The same lower bound applies when the server is not allowed to move before the first request is revealed.
\end{lemma}

\begin{proof}
	Fix an algorithm \ALG in which the server can move only at speed 0 or 1.
	Then there exists some time $\epsilon>0$ such that, before time $\epsilon$, the server has not changed its speed.
	Hence~${\pos(\epsilon)\in \{0,\epsilon\}}$.
	If $\pos(\epsilon)=0$, we release the request $r=(\epsilon,\epsilon)$, and if $\pos(\epsilon)=\epsilon$, we release the request~$r'=(\epsilon,0)$.
	In both cases, we have $\OPT=\epsilon$, while \ALG cannot complete the request before time~$2\epsilon$.
	Therefore, the competitive ratio is at least $2$.
	Observe that even if the server is allowed to move at intermediate speeds, as long as it does not move before the first request is revealed, the same argument still applies.
\end{proof}

Next, we give an algorithm whose competitive ratio matches this lower bound.

\subsection{A 2-competitive algorithm}

\begin{algorithm}[t]
	\caption{$\leftfirst$}\label{alg:leftfirst}
	\uIf{there is a left-request}{
		\movel\;
	}
	\Else{
		\uIf{there is a right-request}{
			\mover\;
		}
		\Else{\textsc{Stay}\;}
	}
\end{algorithm}

When considering algorithms for online TSP on the half-line (or even on the line), observe that, at any point in time, the algorithm only needs to consider at most two requests:
the request with the leftmost position among all unserved requests to the algorithm's left (referred to as the \emph{left-request}) and the request with the rightmost position among all unserved requests to the algorithm's right (referred to as the \emph{right-request}).
At some times, there may be no left-request or no right-request. This is the case when all unserved requests lie on the same side of the server.
If there are multiple requests at the same position, we select the one with the latest release time.
By serving the left- and right-request, the algorithm necessarily also serves all other requests on the way.
In our analysis, we therefore consider only left- and right-requests.
The online algorithm is thus defined by a strategy that determines when to serve the left-request, when to serve the right-request, and when to wait.

The algorithm \leftfirst is formally defined in \Cref{alg:leftfirst} and follows a simple strategy: It prioritizes left-requests over right-requests.
More precisely, whenever a left-request is available, the server executes command \movel, i.e., travels towards the origin at full speed (note that \movel is never invoked when the server is located at the origin).
If no left-request is available but a right-request exists, the server executes command \mover, i.e., it moves away from the origin at full speed.
If there are no unserved requests, the server follows command \textsc{Stay}, i.e., remains stationary.

Before turning to the analysis of the algorithm, we state a simple observation that will be useful throughout our paper.
Since the off\-line optimum cannot serve requests before they are released, the following holds.

\begin{observation}
	\label{obs:opt-t}
	If $t$ is the release time of any request of the input sequence, then $\OPT\geq t$.
\end{observation}

Now we have all the prerequisites at hand to analyze the algorithm \leftfirst.

\begin{theorem}
	\label{thm:leftfirst}
	\leftfirst is 2-competitive.
\end{theorem}

\begin{proof}
	Fix an arbitrary request sequence.
	Let $r=(t,p)$ denote the last request served by~\leftfirst.
	First, observe that, if $r$ is a left-request, we have
	\begin{equation*}
		\leftfirst \leq t+ (\pos(t)-p) \leq t + \pos(t) \leq 2t \overset{\text{Obs.~\ref{obs:opt-t}}}{\leq} 2 \OPT,
	\end{equation*}
	where we have made use of the fact that $\pos(t)\leq t$ since the server cannot move at speed more than 1.
	Therefore, we can assume from now on that $r$ is a right-request.

	Next, observe that, if the server does not move left after time $t$, we have
	\begin{equation*}
		\leftfirst \leq t + (p - \pos(t)) \leq t + p \leq 2t \overset{\text{Obs.~\ref{obs:opt-t}}}{\leq} 2 \OPT,
	\end{equation*}
	where we have used the assumption that every request fulfills $t\geq p$.

	Therefore, we can assume from now on that there was an unserved left-request after time~$t$ and let $r_L=(t_L, p_L)$ denote the last left-request served by \leftfirst.
	Observe that we have
	\begin{equation}
		\label{eq:leftfirst_upper}
		\leftfirst \leq t_L+ (\pos(t_L)-p_L)+(p-p_L).
	\end{equation}
	We now distinguish the following two cases: (1) the off\-line optimum serves $r_L$ before $r$, or (2) the off\-line optimum serves $r$ before $r_L$.
	In the first case, we have
	\begin{equation}
		\label{eq:leftfirst_opt_firstcase}
		\OPT \geq t_L + (p-p_L).
	\end{equation}
	With \eqref{eq:leftfirst_upper} and \eqref{eq:leftfirst_opt_firstcase}, we obtain
	\begin{equation*}
		\leftfirst \leq t_L+ (\pos(t_L)-p_L)+(p-p_L) = \underbrace{t_L+ p-p_L}_{\leq \OPT}+\underbrace{\pos(t_L)}_{\leq t_L\leq \OPT} - \underbrace{p_L}_{\geq 0} \leq 2 \OPT.
	\end{equation*}
	In the second case, i.e., where \OPT serves $r$ before $r_L$, we have
	\begin{equation}
		\label{eq:leftfirst_opt_secondcase}
		\OPT\geq t + (p-p_L).
	\end{equation}
	Moreover, observe that we have $\pos(t_L)\leq t$ because either $t\geq t_L\geq \pos(t_L)$ or $t_L\geq t$ and therefore $r=(t, p)$ was an unserved right-request at time $t_L$.
	With \eqref{eq:leftfirst_upper} and \eqref{eq:leftfirst_opt_secondcase}, we obtain
	\begin{equation*}
		\leftfirst \leq t_L + \underbrace{\pos(t_L)}_{\leq t} - \underbrace{p_L}_{\geq 0} + p - p_L
		\leq \underbrace{t_L}_{\leq \OPT} + \underbrace{t+p-p_L}_{\leq \OPT}
		\leq 2 \OPT.
	\end{equation*}
	To summarize, we have in either case that $\leftfirst\leq 2\OPT$, which completes the proof of the theorem.
\end{proof}

Note that \Cref{thm:leftfirst} and \Cref{lem:2-lower} together complete the proof of \Cref{thm:2-competitive}.

\section{Improved algorithm for open online TSP on the half-line}

In this section, we prove our main result (\Cref{thm:mainthm}), i.e., we prove that the competitive ratio of open online~TSP on the half-line is at most 1.75.
For this, we begin by introducing the algorithm \ouralg.

\subsection{The algorithm \ouralg}

The algorithm \ouralg is formally defined in \Cref{alg:ouralgorithm}.
It is parameterized by a value~$\alpha\in (0,1]$ and we let~$t$ denote the current time.
Its intuition can be summarized as follows:
First, recall that no algorithm is better than 2-competitive if the server does not move while being idle (\Cref{lem:2-lower}), so we let the server move at speed~$\alpha$ before the first request is revealed.
As before, we always consider at most two requests: the left-request and the right-request.
The most important property of our algorithm is that every left-request $r_L=(t_L,p_L)$ has a strict deadline, meaning that it must be served no later than at time $(1+\alpha)t_L$.
To make this possible, the server maintains at every time $t$ that $\pos(t)\leq \alpha t$.
Therefore, whenever a new left-request $r_L=(t_L,p_L)$ appears, the server is located at $\pos(t_L)\leq \alpha t_L$ so that, by moving left immediately, position $p_L\ge 0$ can be reached by time~$(1+\alpha)t_L$.
However, if the server can serve a right-request without violating the deadlines of left-requests, it proceeds to do so.
Note that a right-request $r_R=(t_R, p_R)$ can be reached by time $\max\{t+p_R-\pos(t), \frac{p_R}{\alpha}\}$, where the first term corresponds to moving from~$\pos(t)$ to~$p_R$ at unit speed, and the second term is the arrival time at $p_R$ in case~$\pos(t)\leq \alpha t$ becomes tight.
Therefore, the server prioritizes a right-request over a left-request if and only if $\max\{t+p_R-\pos(t), \frac{p_R}{\alpha}\}+p_R-p_L\leq (1+\alpha)t_L$.

The algorithm definition contains the subroutines \movel and $\mover(\alpha)$.
When \movel is invoked, it simply orders the server to move towards the origin at unit speed (observe that \movel is never invoked when the server is located at the origin).
$\mover(\alpha)$ depends on the parameter $\alpha$ and orders the server to move away from the origin at full speed while maintaining that $\pos(t)\leq \alpha t$. In other words, the server moves at speed~1 as long as~$\pos(t)<\alpha t$ and at speed $\alpha$ when $\pos(t)=\alpha t$.
In particular, the server always moves either at speed 1 or at speed $\alpha$. When it moves at speed $\alpha$, we say that it \emph{slows down}.

\begin{figure}
	\begin{minipage}[t]{0.63\linewidth}
		\begin{algorithm}[H]
			\caption{$\ouralg(\alpha)$}\label{alg:ouralgorithm}
			\uIf{there is a left-request $r_L=(t_L, p_L)$}{
				\uIf{there is a right-request $r_R=(t_R, p_R)$
					with $\max\{t+p_R-\pos(t), \frac{p_R}{\alpha}\}+p_R-p_L \leq (1+\alpha)t_L$
				}{
					$\mover(\alpha)$\;
				}
				\Else{\movel\;}
			}\Else{
				$\mover(\alpha)$\;
			}
		\end{algorithm}
	\end{minipage}
	\hfill
	\begin{minipage}[t]{0.34\linewidth}
		{\renewcommand{\thealgocf}{}
			\begin{algorithm}[H]
				\caption{$\mover(\alpha)$}
				\uIf{$\pos(t)<\alpha t$}{
					move right at speed 1\;
				}
				\Else{
					move right at speed $\alpha$\;
				}
			\end{algorithm}}
		\vspace{.8em}
		{\renewcommand{\thealgocf}{}
			\begin{algorithm*}[H]
				\caption{$\movel$}
				move left at speed 1\;
			\end{algorithm*}}
	\end{minipage}
\end{figure}

Let us summarize the most important properties that immediately follow from the algorithm's definition.

\begin{observation}
	\label{obs:ouralg_properties}
	$\ouralg(\alpha)$ fulfills the following properties.
	\begin{enumerate}[label=\roman*), topsep=0pt]
		\item At every time $t$, the server position fulfills $\pos(t)\leq \alpha t$. \label{obs:pos-alpha}
		\item Every left-request $r_L=(t_L,p_L)$ is served no later than at time $(1+\alpha)t_L$, unless a new left-request appeared before time $(1+\alpha)t_L$. \label{obs:leftrequests}
		\item Every right-request $r_R=(t_R, p_R)$ is served no earlier than at time $p_R/\alpha$. Moreover, if the server is slowed down (i.e., moving at speed $\alpha$) at the time of serving $r_R$, it is served precisely at time $p_R/\alpha$.\label{obs:rightrequests}
	\end{enumerate}
\end{observation}

\subsection{Upper bound for the competitive ratio of \ouralg}
\label{sec:upper-bound-analysis}

We now turn to analyzing the algorithm \ouralg. More precisely, we show the following.

\begin{theorem}
	\label{thm:upperbound_alpha}
	For $\alpha \in (0.3,1]$, the competitive ratio of $\ouralg(\alpha)$ is upper-bounded by
	$\max\{1+\alpha, 1/\alpha, 2.5-\alpha, 1+1/(2\alpha)\}$.
\end{theorem}

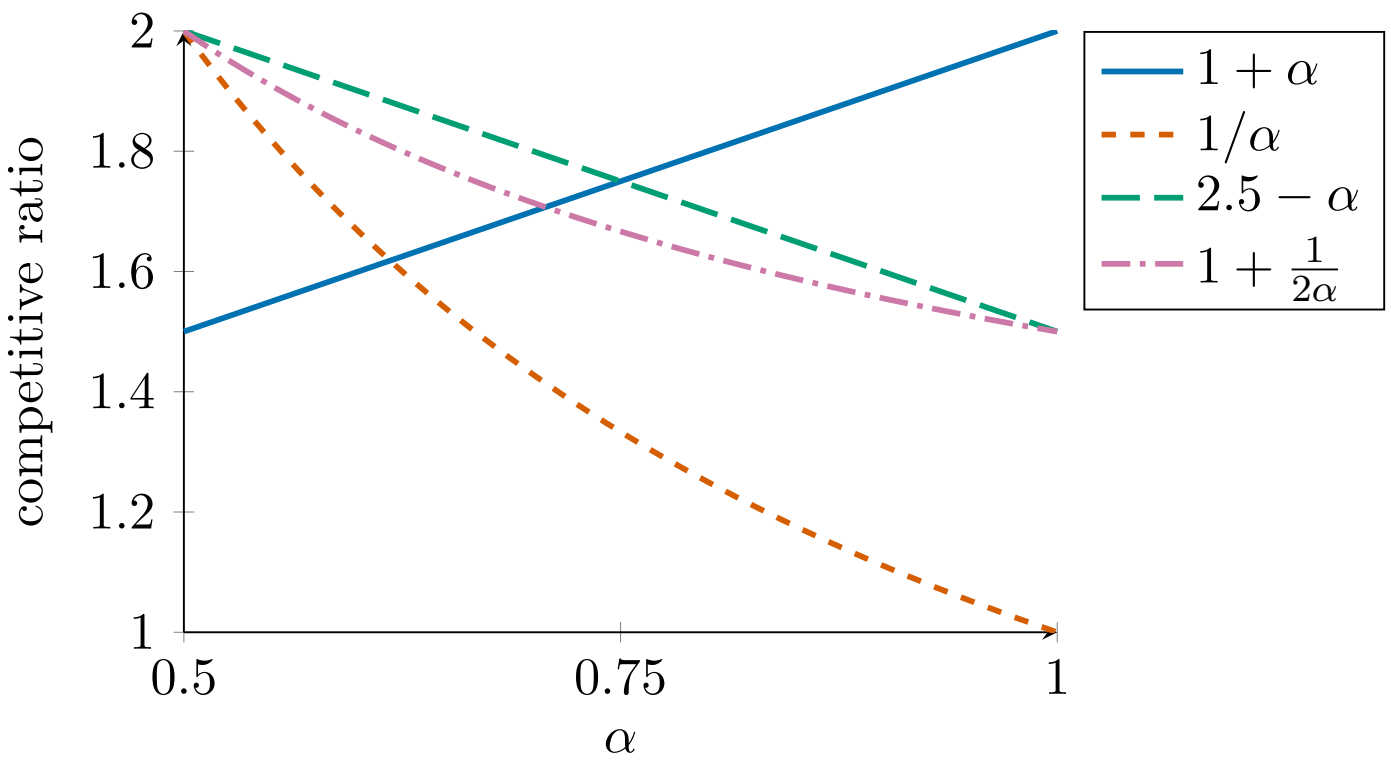
\begin{figure}
	\centering
		\begin{tikzpicture}[scale=0.9]
			\begin{axis}
				[
					axis lines = left,
					xlabel = {$\alpha$},
					ylabel = {competitive ratio},
					ymin=1, ymax=2,
					xmin=0.5, xmax=1,
					domain=0.5:1,
					samples=100,
					legend pos=outer north east,
					legend cell align={left},
					height=6cm,
					width=8cm,
					xtick={0.5, 0.75, 1},
				]
				\addplot [very thick, blue] {1 + x};
				\addlegendentry{$1+\alpha$}
				\addplot [very thick, orange, dashed] {1/x};
				\addlegendentry{$1/\alpha$}
				\addplot [very thick, green, dash pattern=on 8pt off 3pt] {2.5 - x};
				\addlegendentry{$2.5-\alpha$}
				\addplot [very thick, rose, dash pattern=on 6pt off 2pt on \pgflinewidth off 2pt] {1 + 1/(2*x)};
				\addlegendentry{$1 + \frac{1}{2\alpha}$}
			\end{axis}
		\end{tikzpicture}
	\caption{Bounds on the competitive ratio from \Cref{thm:upperbound_alpha} as a function of~$\alpha$.}
	\label{fig:bounds}
\end{figure}

Note that, by setting $\alpha=0.75$, we obtain a competitive ratio of 1.75 (cf.~\Cref{fig:bounds}), i.e., the statement of the first part of \Cref{thm:mainthm} follows.

We now prove \Cref{thm:upperbound_alpha}.
To this end, for the remainder of this subsection, we fix an arbitrary input sequence of requests, and we show that $\ouralg(\alpha)$ achieves the claimed ratio on this sequence.
We denote by $\OPT$ an optimum off\-line solution, or its completion time, depending on context.
We denote the last right-request served by $\ouralg(\alpha)$ by~${r_R=(t_R, p_R)}$, and the last left-request served by $\ouralg(\alpha)$ by~${r_L=(t_L, p_L)}$.
In case there is no left-request, Observations~\ref{obs:opt-t} and~\ref{obs:ouralg_properties}~\ref{obs:rightrequests} immediately bound the competitive ratio by $1/\alpha$, and in case there is no right-request, Observations~\ref{obs:opt-t} and~\ref{obs:ouralg_properties}~\ref{obs:leftrequests} bound it by $1+\alpha$.
Similarly, we immediately obtain the desired bounds in the following two cases.

\begin{observation}
	\label{obs:left_last}
	If one of the following conditions holds, then $\ouralg(\alpha)$ completes the request sequence by $\max(1+\alpha, 1/\alpha)\OPT$:
	\begin{enumerate}[label=\roman*)]
		\item $\ouralg(\alpha)$ serves $r_L$ after $r_R$.
		\item $\ouralg(\alpha)$ serves $r_R$ while moving at speed~$\alpha$.
	\end{enumerate}
\end{observation}

Therefore, we assume from now on that~$r_L$ and $r_R$ both exist, and that the server first served~$r_L$ and then moved at full speed to $r_R$.
Next, we identify two further simple cases within these assumptions.

\begin{lemma}
	\label{lem:opt_serves_r_first}
	Assume that $\ouralg(\alpha)$ serves $r_L$ before $r_R$ and always moves at unit speed after serving $r_L$.
	If one of the following two conditions holds, the request sequence is completed by time $(1+\alpha)\OPT$.
	\begin{enumerate}[label=\roman*)]
		\item \OPT serves $r_L$ before $r_R$.
		\item $t_L\leq t_R$.
	\end{enumerate}
\end{lemma}
\begin{proof}
	In both cases, \OPT serves neither of $r_L, r_R$ earlier than at time $t_L$, and then travels to the other request, which takes $p_R-p_L$ time units. Therefore,
	\begin{equation}
		\label{eq:opt-case-2-2}
		\OPT \geq t_L+ p_R-p_L\,.
	\end{equation}

	By assumption, \ouralg first serves $r_L$ and then moves at full speed to $p_R$.
	By Observation~\ref{obs:ouralg_properties}~\ref{obs:leftrequests}, $r_L$ is reached no later than $(1+\alpha)t_L$ and then moving to $r_R$ takes $p_R-p_L$ time units.
	It follows that the request sequence is completed by time
	\[
		(1+\alpha)t_L+p_R-p_L = t_L+p_R-p_L + \alpha t_L
		\overset{\eqref{eq:opt-case-2-2}}{\leq} \OPT + \alpha t_L
		\overset{\text{Obs.~\ref{obs:opt-t}}}{\leq}(1+\alpha)\OPT\,.\qedhere
	\]
\end{proof}

The remaining case turns out to be the most difficult to analyze and constitutes the crux of our proof.
While the two previous results hold for any $\alpha\in (0,1]$, we need in the following lemma that $\alpha\geq 0.3$.
This is a reasonable assumption, since for $\alpha\leq 0.3$, the bound $1/\alpha$ from \Cref{obs:left_last} is already significantly worse than the bound of 2 achieved by \leftfirst.

\begin{lemma}
	\label{lem:final_case}
	Let $\alpha\geq 0.3$. Assume that $\ouralg(\alpha)$ serves $r_L$ before $r_R$ and always moves at unit speed after serving $r_L$. Additionally, assume that $t_L>t_R$ and \OPT serves $r_R$ before $r_L$.
	Then $\ouralg(\alpha)$ completes the request sequence by time $\max(2.5-\alpha, 1+1/(2\alpha))\OPT$.
\end{lemma}
\begin{proof}
	First, observe that in the given setting, when $r_L$ appeared at time $t_L$, the server decided not to detour to serve $r_R$ before $r_L$. In other words, at time $t_L$, the condition in line~2 of \Cref{alg:ouralgorithm} was violated, i.e., we have
	\begin{equation}
		\label{eq:detour-violation}
		\max\left\{t_L+p_R-\pos(t_L), \frac{p_R}{\alpha}\right\}+(p_R-p_L) > (1+\alpha)t_L\,.
	\end{equation}
	Further, note that the completion time of $\ouralg(\alpha)$ is given by
	\begin{equation}
		\label{eq:algs_completion_time}
		\ouralg(\alpha) \leq t_L + (\pos(t_L)-p_L)+ (p_R-p_L) = t_L + \pos(t_L) + p_R - 2 p_L
	\end{equation}
	and the completion time of the off\-line optimum is bounded by
	\begin{equation}
		\label{eq:opts_completion_time}
		\OPT\geq t_R+p_R-p_L.
	\end{equation}
	Next, we will make use of the following observation: The behavior of \ouralg is invariant under multiplying all release times and positions by the same factor $x>0$.
	In particular, this allows us to assume w.l.o.g.~that $\OPT=1$ by rescaling by $1/\OPT$
	(if $\OPT=0$, then $\ALG=0$ is trivial).
	Summarizing our findings, it follows that the competitive ratio is bounded by the maximum value of the following optimization problem (here, \ALG, \OPT, $t_R, t_L, p_R, p_L, \pos(t_L)$ are interpreted as variables of the optimization problem):
	\begin{align*}
		\max\quad        & \ALG                                                                                                     \\
		\text{s.t.}\quad & \OPT \geq t_L\tag{\Cref{obs:opt-t}}                                                                      \\
		                 & \OPT\geq t_R+p_R-p_L\tag{\ref{eq:opts_completion_time}}                                                  \\
		                 & \ALG \leq t_L + \pos(t_L) + p_R-2p_L\tag{\ref{eq:algs_completion_time}}                                  \\
		                 & \max(t_L + p_R-\pos(t_L), p_R/\alpha)+p_R-p_L \geq (1+\alpha)t_L\tag{\ref{eq:detour-violation}, relaxed} \\
		                 & \pos(t_L)\leq p_R \tag{$r_R$ is a right-request}                                                         \\
		                 & t_R\geq p_R\tag{by assumption}                                                                           \\
		                 & \OPT = 1\tag{scaling}                                                                                    \\
		                 & \ALG, \OPT, t_R, t_L, p_R, p_L, \pos(t_L) \geq 0\,.
	\end{align*}
	Observe that the problem is not linear due to the maximum in \eqref{eq:detour-violation}.
	Therefore, we now distinguish the two cases $t_L + p_R-\pos(t_L) \geq p_R/\alpha$ and $t_L + p_R-\pos(t_L) < p_R/\alpha$.
	For each of the two cases, we obtain a linear program, which allows us to use duality.
	In the end, the competitive ratio will be bounded by the maximum of the two values of the optimization problems.

	First, we consider the case $t_L + p_R-\pos(t_L) \geq p_R/\alpha$.
	Here, we obtain the following LP, where it turns out that the condition $\pos(t_L)\leq p_R$ can be dropped.
	\begin{align*}
		\max\quad        & \ALG                                                                                   \\
		\text{s.t.}\quad & \OPT \geq t_L\tag{\Cref{obs:opt-t}}                                                    \\
		                 & \OPT\geq t_R+p_R-p_L\tag{\ref{eq:opts_completion_time}}                                \\
		                 & \ALG \leq t_L + \pos(t_L) + p_R-2p_L\tag{\ref{eq:algs_completion_time}}                \\
		                 & t_L + p_R-\pos(t_L)+p_R-p_L \geq (1+\alpha)t_L\tag{\ref{eq:detour-violation}, relaxed} \\
		                 & t_R\geq p_R\tag{by assumption}                                                         \\
		                 & \OPT = 1\tag{scaling}                                                                  \\
		                 & \ALG, \OPT, t_R, t_L, p_R, p_L, \pos(t_L) \geq 0\,.
	\end{align*}
	Writing $\x=(\ALG,\OPT,t_R,t_L,p_R,p_L,\pos(t_L))^\top$, a relaxation $(\mathrm{P})$ of the LP and its dual $(\mathrm{D})$ can equivalently be written as
	\begin{align*}
		\max\quad(1,0,0,0,0,0,0)^\top & \x\quad \text{s.t.} \\
		(\mathrm{P})\quad\begin{pmatrix}
			0 & -1 & 0  & 1      & 0  & 0  & 0  \\
			0 & -1 & 1  & 0      & 1  & -1 & 0  \\
			1 & 0  & 0  & -1     & -1 & 2  & -1 \\
			0 & 0  & 0  & \alpha & -2 & 1  & 1  \\
			0 & 0  & -1 & 0      & 1  & 0  & 0  \\
			0 & 1  & 0  & 0      & 0  & 0  & 0
		\end{pmatrix} & \x\leq\begin{pmatrix}
			0 \\ 0 \\ 0 \\ 0 \\ 0 \\ 1
		\end{pmatrix} \\
		& \x\geq \boldsymbol{0}
	\end{align*}
	and
	\begin{align*}
		\min\quad (0,0,0,0,0,1)^\top & \y\quad \text{s.t.} \\
		(\mathrm{D})\quad\begin{pmatrix}
			0  & 0  & 1  & 0      & 0  & 0 \\
			-1 & -1 & 0  & 0      & 0  & 1 \\
			0  & 1  & 0  & 0      & -1 & 0 \\
			1  & 0  & -1 & \alpha & 0  & 0 \\
			0  & 1  & -1 & -2     & 1  & 0 \\
			0  & -1 & 2  & 1      & 0  & 0 \\
			0  & 0  & -1 & 1      & 0  & 0
		\end{pmatrix} & \y\geq\begin{pmatrix}
			1 \\ 0 \\ 0 \\ 0 \\ 0 \\ 0 \\ 0
		\end{pmatrix} \\
		& \y\geq \boldsymbol{0}\,.
	\end{align*}
	By weak duality, any feasible solution to the dual provides an upper bound on the optimum value of the primal.
	The dual solution $\y=(1-\alpha,1.5,1,1,1.5,2.5-\alpha)^\top$ is feasible for~$\alpha \in (0, 1]$ and has objective value $2.5-\alpha$.

	Next, we consider the case $t_L + p_R-\pos(t_L) < p_R/\alpha$.
	Here, we obtain the following LP, where it turns out that the condition $\OPT\geq t_L$ can be dropped.
	\begin{align*}
		\max\quad        & \ALG                                                                                  \\
		\text{s.t.}\quad & \OPT\geq t_R+p_R-p_L\tag{\ref{eq:opts_completion_time}}                               \\
		                 & \ALG \leq t_L + \pos(t_L) + p_R-2p_L\tag{\ref{eq:algs_completion_time}}               \\
		                 & \frac{1}{\alpha} p_R+p_R-p_L \geq (1+\alpha)t_L\tag{\ref{eq:detour-violation}, relaxed} \\
		                 & \pos(t_L)\leq p_R \tag{$r_R$ is a right-request}                                      \\
		                 & t_R\geq p_R\tag{by assumption}                                                        \\
		                 & \OPT = 1\tag{scaling}                                                                 \\
		                 & \ALG, \OPT, t_R, t_L, p_R, p_L, \pos(t_L) \geq 0\,.
	\end{align*}
	As before, we can write the dual $(\mathrm{D})$ of the relaxed LP as
	\begin{align*}
		\min\quad (0,0,0,0,0,1)^\top & \y\quad \text{s.t.} \\
		(\mathrm{D})\quad\begin{pmatrix}
			0  & 1  & 0                        & 0  & 0  & 0 \\
			-1 & 0  & 0                        & 0  & 0  & 1 \\
			1  & 0  & 0                        & 0  & -1 & 0 \\
			0  & -1 & 1+\alpha                 & 0  & 0  & 0 \\
			1  & -1 & -\frac{\alpha+1}{\alpha} & -1 & 1  & 0 \\
			-1 & 2  & 1                        & 0  & 0  & 0 \\
			0  & -1 & 0                        & 1  & 0  & 0
		\end{pmatrix} & \y\geq\begin{pmatrix}
			1 \\ 0 \\ 0 \\ 0 \\ 0 \\ 0 \\ 0
		\end{pmatrix} \\
		& \y\geq \boldsymbol{0}\,.
	\end{align*}
	The dual solution $\y=\bigl(1+\frac{1}{2\alpha},1,\frac{1}{1+\alpha},1,1+\frac{1}{2\alpha},1+\frac{1}{2\alpha}\bigr)^\top$ is feasible for $\alpha \geq 0.3$
	(more precisely, the second-to-last inequality holds for $\alpha\geq (\sqrt{17}-3)/4\approx0.281$, and all other inequalities hold for all $\alpha>0$).
	The corresponding objective value is $1+\frac{1}{2\alpha}$.

	Together, we obtain that the competitive ratio is bounded by $\max(2.5-\alpha, 1+1/(2\alpha))$.
\end{proof}

Let us summarize the different cases we considered to prove \Cref{thm:upperbound_alpha}.

\begin{proof}[Proof of \Cref{thm:upperbound_alpha}]
	First, we noted that, in case there is no left-request or no right-request, we easily obtain a bound of $\max(1+\alpha, 1/\alpha)$ on the competitive ratio.
	Therefore, we can assume that a left- and right-request exist and let $r_L$ denote the last-served left-request and~$r_R$ the last-served right-request.
	By \Cref{obs:left_last}, we can assume that the server first served~$r_L$ and then moved at full speed to $r_R$ (otherwise, we obtain a bound of $\max(1+\alpha, 1/\alpha)$).
	By \Cref{lem:opt_serves_r_first}, we can additionally assume that \OPT serves $r_R$ before $r_L$ and that $r_R$ was released before~$r_L$ (otherwise, we obtain a bound of $1+\alpha$).
	Finally, we showed in \Cref{lem:final_case} that, under these assumptions, the competitive ratio is bounded by $\max(2.5-\alpha, 1+1/(2\alpha))$.
\end{proof}

\subsection{Lower bound for the competitive ratio of \ouralg}
\label{sec:alg-lower-bounds}
In this subsection, we present two lower-bound constructions for \ouralg, showing that the analysis in \Cref{sec:upper-bound-analysis} is tight in the sense that no choice of $\alpha$ yields a competitive ratio below~$1.75$ for $\ouralg(\alpha)$ (i.e., we prove the second part of \Cref{thm:mainthm}).

\begin{lemma}
	\label{lem:lower_bound_ouralg_firstcase}
	The competitive ratio of $\ouralg(\alpha)$ is at least $\max(1+\alpha, 1/\alpha)$ for every~${\alpha \in (0, 1]}$.
\end{lemma}
\begin{proof}
	First, consider the input sequence consisting of the single request $(1,0)$.
	Note that $\ouralg(\alpha)$ executes $\mover(\alpha)$ before time 1 so that $\pos(1)=\alpha$.
	Therefore, the request is served by time $1+\alpha$.
	On the other hand, \OPT serves the request immediately at time~$1$ by waiting at the origin.
	Therefore, the competitive ratio is at least $1+\alpha$.

	Next, consider the input sequence consisting of the single request $(1,1)$.
	Since $(1,1)$ is a right-request, by \Cref{obs:ouralg_properties}~\ref{obs:rightrequests}, $\ouralg(\alpha)$ completes it not earlier than at time~$1/\alpha$, while we have $\OPT=1$. Therefore, the competitive ratio is at least $1/\alpha$.
\end{proof}

\begin{lemma}
	\label{lem:lower_bound_ouralg_secondcase}
	The competitive ratio of $\ouralg(\alpha)$ is at least $2.5-\alpha$ for $\alpha \in [0.5, 1]$.
\end{lemma}
\begin{proof}
	Consider the input sequence consisting of the three requests $(\frac{\alpha}{1+\alpha}, 0)$, $(0.5+\epsilon, 0.5+\epsilon)$ and~$(1, 0)$ where $0 < \epsilon<0.5$.
	Observe that $\alpha/(1+\alpha)\leq 0.5$ for $\alpha\leq 1$, so
	the requests indeed arrive in the order given above.

	\OPT can serve the requests by moving to position $0.5+\epsilon$ at time $0.5+\epsilon$ and then moving to position $0$, i.e., we have
	\begin{equation}
		\label{eq:opt_lower_bound_for_ouralg}
		\OPT=1+2\epsilon.
	\end{equation}

	On the other hand, $\ouralg(\alpha)$ moves away from the origin at speed $\alpha$ until time~${\alpha/(1+\alpha)}$ so that $\pos(\alpha/(1+\alpha))=\alpha^2/(1+\alpha)$.
	By moving back to the origin at unit speed, $\ouralg(\alpha)$ could serve the first request at time
	\begin{equation*}
		\frac{\alpha}{1+\alpha} + \frac{\alpha^2}{1+\alpha} = \alpha = (1+\alpha)\cdot \frac{\alpha}{1+\alpha}\,,
	\end{equation*}
	which is the maximum time allowed to serve the first request according to the algorithm definition.
	Therefore, $\ouralg(\alpha)$ cannot decide to serve another right-request before serving the left-request at position $0$, or before another left-request is revealed.
	Hence, $\ouralg(\alpha)$ serves the left-request at position $0$ at time $\alpha\leq1$, and then moves right at maximum speed while maintaining $\pos(t)\leq \alpha t$ to serve the right-request at position $0.5+\epsilon$.
	Since $\alpha\geq 0.5$, we obtain $\pos(1)=1-\alpha<0.5+\epsilon$, i.e., the request $(0.5+\epsilon, 0.5+\epsilon)$ is still unserved when the new left-request $(1,0)$ arrives.

	If $\ouralg(\alpha)$ served the right-request first, it would arrive at the left-request at time
	\begin{equation*}
		1+(0.5+\epsilon)-\pos(1)+0.5+\epsilon=2+2\epsilon-(1-\alpha)
		= 1+\alpha+2\epsilon > (1+\alpha)\cdot 1\,,
	\end{equation*}
	hence $\ouralg(\alpha)$ has to move to the left-request first before serving the right-request.
	It follows that $\ouralg(\alpha)$ finishes at time
	\begin{equation*}
		1+\pos(1)+0.5+\epsilon=
		1+(1-\alpha)+0.5+\epsilon = 2.5 - \alpha + \epsilon\,.
	\end{equation*}
	Together with \eqref{eq:opt_lower_bound_for_ouralg} and letting $\epsilon \to 0$, we obtain that the competitive ratio is at least~$2.5-\alpha$.
\end{proof}

For $\alpha\leq 0.5$, \Cref{lem:lower_bound_ouralg_firstcase} yields the lower bound~$1/\alpha \geq 2$ on the competitive ratio of $\ouralg(\alpha)$.
For $\alpha >0.5$, the competitive ratio is at least $\max(1+\alpha, 2.5-\alpha)\geq 1.75$ by \Cref{lem:lower_bound_ouralg_firstcase,lem:lower_bound_ouralg_secondcase}.
This completes the proof that $\ouralg(\alpha)$ has competitive ratio at least 1.75 for any choice of $\alpha$ (i.e., the second part of \Cref{thm:mainthm}).

\section{General lower bound for online TSP on the half-line}
\label{sec:lower-improved}

In the following, we fix an arbitrary online algorithm \ALG and prove a general lower bound for the competitive ratio of online TSP on the half-line.

\thmLowerBound*

It is immediate to see that no online algorithm for the TSP on the half-line can achieve a competitive ratio strictly smaller than $3/2$:
At time $1$, we reveal either a request at $0$ if~${\pos(1)\geq 1/2}$, or a request at $1$ otherwise.
In both cases, we have $\ALG\geq 3/2$ and $\OPT=1$.

For the remainder of this section, we fix $\rho\in (3/2, 5/3]$ and assume that \ALG has competitive ratio strictly less than $\rho$.
We progressively introduce requests that, based on the fact that it must be~$\rho$-competitive if no further requests appear, constrain \ALG's behavior more and more.
Specifically, we present two fixed requests, followed by three requests that depend on \ALG's behavior.
Based on the derived restrictions on \ALG, we show that the resulting request sequence yields a lower bound of~$\rho \geq 1.6463$.
We denote by $\servingTime{i}$ the time when the request~$r_i=(t_i, p_i)$ is served by \ALG and formally define the adversarial request sequence as follows.

\begin{definition}[Adversarial request sequence]
	\label{def:adversary}
	We present an adversarial request sequence $(r_1,r_2,r_3,r_4,r_5)$ consisting of the requests
	\begin{enumerate}
		\item $r_1\coloneq(1/\rho, 0)$ and $r_2\coloneq(1,1)$,
		\item $r_3\coloneq(t^*,0)$, where $t^*\coloneq \min\{t\geq 1:\pos(t)=2\rho-1-t\}$,
		\item $r_4\coloneq(b,b)$ for some $b\geq t^*+1/2$ to be determined later,
		\item $r_5$ depending on \ALG's behavior:
		      \begin{enumerate}[label=\emph{\roman*.}]
			      \item if $\pos(2b)<b/2$ and $\servingTime{4}> 2b$, then $r_{5}\coloneq(t_{\mathrm{m}},0)$ with $t_{\mathrm{m}}\coloneq\min\{t\geq 2b:\pos(t)=b/2\}$,
			      \item otherwise, $r_{5}\coloneq(2b, 0)$.
		      \end{enumerate}
	\end{enumerate}
\end{definition}

We show in \Cref{lem:line_intersection,lem:lower_bound_firstcase} that this sequence is well-defined, in the sense that the time~$t^*$ in the second step and, if needed, the time~$t_{\mathrm{m}}$ in the fourth step exist.

\begin{figure}
	\centering
	\newcommand{\xtick}[2]{\draw (#1,0.05) -- (#1,-0.05); \node at (#1,-0.2) {\footnotesize #2 \vphantom{pb}};}
	\newcommand{\ytick}[2]{\draw (0.05,#1) -- (-0.05,#1); \node at (-0.2,#1) {\footnotesize #2};}
	\begin{tikzpicture}[scale=1.5]

		% instance parameters for rho = 1.64
		\def\rhov{1.64}
		\def\brho{2.18}
		\def\tstar{1.55}
		\def\tm{5.0272}

		\draw[->] (0,0) -- (8.7,0) node[right] {$t$};
		\draw[->] (0,0) -- (0,2.5) node[above] {$\pos(t)$};

		% trajectory of OPT
		\draw[thick, orange] (0,0) -- (\brho, \brho) -- (2*\brho, 0) -- ({\rhov*(\tstar+\brho)-\brho/2}, 0);

		% requests
		\filldraw[black] (1/\rhov, 0) circle (1.3pt);
		\draw (1/\rhov, 0) -- (0.3,-0.12)node[below left,xshift=3pt,yshift=2pt] {\small $r_1$};
		\filldraw[black] (1, 1) circle (1.3pt) node[below,xshift=2pt] {\small $r_2$};
		\filldraw[black] (\tstar, 0) circle (1.3pt) node[above] {\small $r_3$};
		\filldraw[black] (\brho, \brho) circle (1.3pt) node[above] {\small $r_4$};

		\xtick{\brho}{$b$}
		\xtick{1/\rhov}{$1/\rho$}
		\xtick{1}{$1$}
		\xtick{2*\brho}{$2b$}
		\xtick{\tstar}{$t^*$}
		\xtick{{\rhov*(\tstar+\brho)-\brho/2}}{$t_{\mathrm{m}}$}

		\ytick{1}{$1$}
		\ytick{\brho}{$b$}

		% the line 2*rho-1-t
		\draw[dotted, thick] (0, 2*\rhov-1) -- node[pos=0.25, below, sloped] {\small $2\rho-1-t$} (2*\rhov-1, 0);

		% trajectory of ALG up to time t^*
		\draw[thick, blue] (0,0) -- (1/\rhov, 2/\rhov-1) -- (3/\rhov-1, 0) -- (\tstar, 2*\rhov-1-\tstar);

		% case (i)
		\draw[thick, blue] ({\tstar}, {2*\rhov-1-\tstar})
		.. controls ({\tstar+0.2}, {2*\rhov-0.8-\tstar}) and (1.95, 1) .. (2.15, 1)
		.. controls (2.3, 1) and (3.3, 0) ..
		node[pos=0.5, above right, inner sep=1pt] {i}
		(3.6, 0)
		.. controls (3.9, 0) and (4.68, 0.76) ..
		(\tm, {\brho/2});
		\draw[thick, blue, dashed] (\tm, {\brho/2}) -- ({\rhov*(\tstar+\brho)}, 0) -- ({\rhov*(\tstar+\brho)+\brho}, \brho);
		\draw[thick, blue, dashed] (\tm, {\brho/2})
		-- ({\rhov*(\tstar+\brho)}, \brho) -- ({\rhov*(\tstar+\brho)+\brho}, 0);

		\filldraw[black] ({\rhov*(\tstar+\brho)-\brho/2}, 0) circle (1.3pt) node[above] {\small $r_{5}^{(i)}$};

		% case (ii)
		\def\trtwo{2*\tstar+2-2*\rhov}
		\def\trthree{3+2*\tstar-2*\rhov}
		\draw[thick, blue] (\tstar, {2*\rhov-1-\tstar}) -- (\trtwo, 1) -- node[pos=0.4, below left, inner sep=1pt] {ii} (\trthree, 0) -- (\trthree+\brho, \brho) -- ({\trthree+2*\brho}, 0);

		\filldraw[black] ({2*\brho}, 0) circle (1.3pt) node[above, xshift=4pt] {\small $r_{5}^{(ii)}$};

	\end{tikzpicture}
	\caption{Adversarial request sequence for the lower bound in \Cref{thm:lower_bound} for $\rho=1.64$, with the best possible trajectory of \ALG in blue and the trajectory of the off\-line optimum in orange.
	Curved lines in \ALG's trajectory indicate that the online algorithm has degrees of freedom in its behavior, whereas straight lines indicate that any other behavior increases the lower bound on the competitive ratio.
	Dashed lines show that there are two possible trajectories.
	The trajectories marked (i) and (ii) correspond to the two cases in the definition of $r_5$.}
	\label{fig:lower_bound_adversary}
\end{figure}
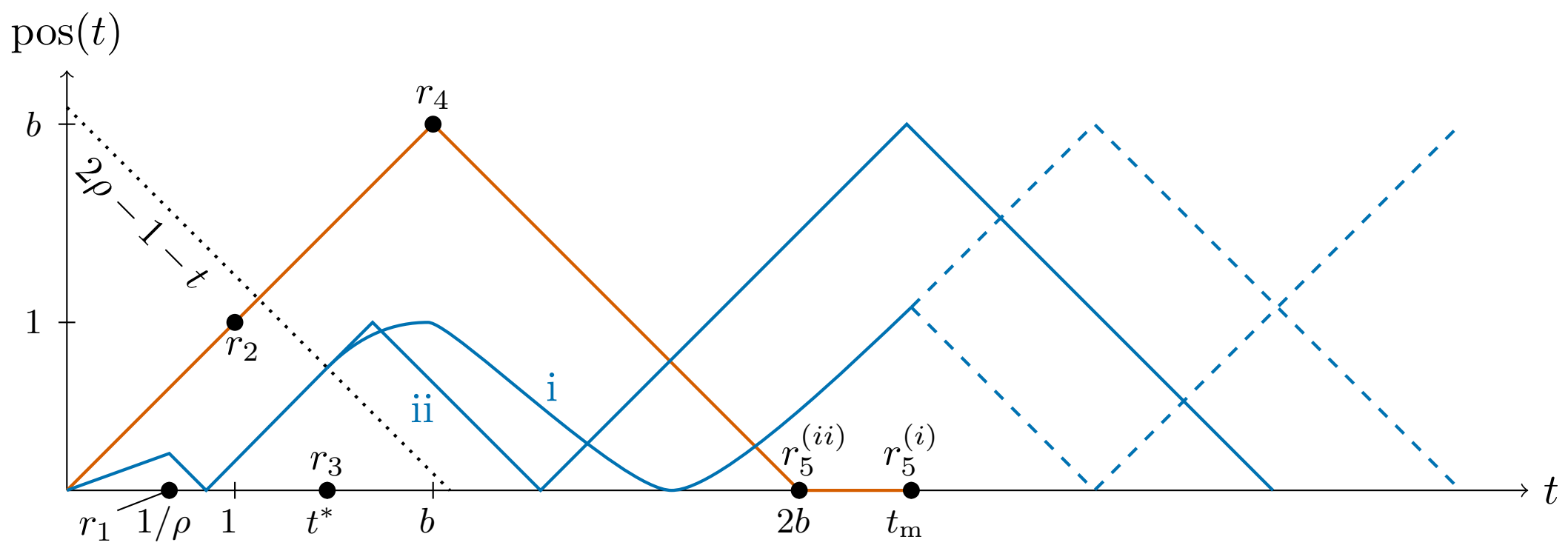

The adversarial sequence is illustrated in \Cref{fig:lower_bound_adversary} for $\rho=1.64$.
The figure also shows the trajectory of the optimum off\-line solution for the entire request sequence, as well as the best possible trajectory of \ALG.

By~$\sigma_{\leq i}\coloneq(r_1,r_2,\dots,r_i)$ we denote the sequence of requests revealed up to and including request $r_i$, and we write $\ALG(\sigma_{\leq i})$ and $\OPT(\sigma_{\leq i})$ for the cost of \ALG and the off\-line optimum, respectively, for serving the request sequence $\sigma_{\leq i}$.
Moreover, we write $\ALG[\sigma_{\leq i}]$ for the schedule \ALG computes for the request sequence $\sigma_{\leq i}$, i.e., \ALG's behavior if only the requests~$r_1,\dots,r_i$ were revealed.
Note that the serving times $\servingTime{i}$ are always w.r.t.~the full schedule $\ALG[\sigma_{\leq 5}]$.

The following lemma establishes that the time $t^*$ as described in the second step of \Cref{def:adversary} exists and falls within certain bounds.

\begin{lemma}
	\label{lem:line_intersection}
	After releasing the requests $r_1$ and $r_2$ of \Cref{def:adversary}, there exists a time~$t^*$ with~${t^*\in[ \rho +(3/(2\rho))-1,2]}$ such that $\pos(t^*)=2\rho-1-t^*$.
	Moreover, $\servingTime{2}>t^*$.
\end{lemma}

\begin{proof}
	Observe that, before the first request $r_1$ is revealed, the server needs to be able to handle the case where the first request is of the form $(t,t)$ with $t\leq 1/\rho$.
	In this case, the off\-line optimum's cost is $t$ and the server must be able to serve the request before time $\rho t$.
	Note that~$\pos(t)\leq t$ since the server is restricted to unit speed.
	It follows that~$t+(t-\pos(t))\leq \rho t$.
	Equivalently, we have
	\begin{equation*}
		\pos(t)\geq (2-\rho)t\quad\text{for all }t\in[0, 1/\rho].
	\end{equation*}

	This implies that, when $r_1$ is revealed at time $1/\rho$, we have $\pos(1/\rho)\geq (2-\rho)/\rho$, and it follows that
	\begin{equation}
		\label{eq:r1_served}
		\servingTime{1}\geq t_1+|\pos(t_1)-p_1|=1/\rho+\pos(1/\rho)\geq (3-\rho)/\rho.
	\end{equation}
	Moreover, $\servingTime{1}\leq 1$ because \ALG is $\rho$-competitive for the request sequence $\sigma_{\leq 1}$, i.e., $r_1$ is already served when $r_2$ is revealed.
	From \eqref{eq:r1_served}, it follows that
	\begin{equation*}
		\servingTime{2}\geq \servingTime{1}+|p_2-\pos(\servingTime{1})|=\servingTime{1}+1\geq (3-\rho)/\rho +1 =3/\rho.
	\end{equation*}
	With this, we obtain that there is a time $t^*$ where $\pos(t^*)=2\rho-1-t^*$:
	Let~$\ell(t)\coloneq 2\rho-1-t$.
	At time $\servingTime{1}$, we have $\pos(\servingTime{1})=0< \ell(\servingTime{1})$.
	When request $r_2=(1,1)$ is served at time~$\servingTime{2}\geq 3/\rho$, we have~$\pos(\servingTime{2})=1 >2\rho-1-(3/\rho)\geq \ell(\servingTime{2})$, where the first inequality holds for $\rho\leq 1.82$.
	Therefore, by continuity of $\pos$ and $\ell$, there exists~$t^*\in (\servingTime{1},\servingTime{2})$ such that~$\pos(t^*)=\ell(t^*)=2\rho-1-t^*$.
	Since $\servingTime{1}\leq 1$ and, as we show below, every time $t\geq \servingTime{1}$ with $\pos(t)=\ell(t)$ satisfies $t\geq \rho+3/(2\rho)-1>1$, the minimum in the definition of~$t^*$ exists and lies in $(\servingTime{1},\servingTime{2})$.

	It remains to prove the claimed upper and lower bounds on $t^*$.
	To this end, note that~$r_2$ is served no earlier than time $t^*+(1-\pos(t^*))=t^*+1-(2\rho -1 -t^*)=2t^*+2-2\rho$.
	Moreover,~$\OPT(\sigma_{\leq 2})\leq 1+1/\rho$ (wait at $p_1=0$ until time $t_1=1/\rho$ and then move up to~$p_2=1$) so that~$2t^*+2-2\rho \leq \rho\OPT(\sigma_{\leq 2})\leq \rho +1$.
	Reformulating and using $\rho \leq 5/3$ yields
	\begin{equation*}
		t^*\leq \frac{3\rho -1}{2} \leq 2\,.
	\end{equation*}
	To bound $t^*$ from below, observe that, using \eqref{eq:r1_served} and $\pos(\servingTime{1})=p_1=0$, we have for all~$t\geq \servingTime{1}$ that $\pos(t)\leq \pos(\servingTime{1})+t-\servingTime{1} \leq t-(3-\rho)/\rho$.
	By definition of time~$t^*$, it therefore holds that
	\begin{equation*}
		2\rho-1-t^*=\pos(t^*) \leq t^*-(3-\rho)/\rho\,,
	\end{equation*}
	or, equivalently, $t^*\geq \rho+3/(2\rho) -1$.
	The same argument applies to every time $t\geq\servingTime{1}$ with $\pos(t)=\ell(t)$.
\end{proof}

To prove restrictions on \ALG's behavior, we use the fact that \ALG's competitive ratio is strictly less than $\rho$ for the request sequences $\sigma_{\leq 3}$ and $\sigma_{\leq 4}$, whose optimum off\-line cost can be bounded.

\begin{observation}
	\label{obs:optimum_cost_after_r3_r4}
	It holds that
	\begin{equation*}
		\OPT(\sigma_{\leq 3})\leq2\quad\text{and}\quad
		\OPT(\sigma_{\leq 4})\leq t^*+b.
	\end{equation*}
\end{observation}
\begin{proof}
	An off\-line algorithm can serve $\sigma_{\leq 3}$ by moving to $p_2=1$ at time $t_2=1$ and then moving back to $p_1=p_3=0$ to serve $r_1=(1/\rho, 0)$ and $r_3=(t^*, 0)$ at time $2$, since $t^*\leq 2$, achieving completion time $2$.

	For $\sigma_{\leq 4}$, the off\-line optimum can wait at~$0$ until time $t^*$ and then move to $p_4=b>1$, serving~$r_2$ on the way.
\end{proof}

The time $t^*$ is chosen precisely such that \ALG is forced to serve $r_2$ before $r_3$, as the alternative would not be competitive in case no more requests are released.

\begin{observation}[Planned order of $r_2$ and $r_3$]
	\label{obs:planned_order_r2_r3}
	$\ALG[\sigma_{\leq 3}]$ serves $r_2=(1,1)$ before $r_3=(t^*, 0)$.
\end{observation}
\begin{proof}
	Otherwise, $\ALG[\sigma_{\leq 3}]$ would finish the request sequence no earlier than
	\begin{equation*}
		t^*+(\pos(t^*)-0)+1=t^*+2\rho-1-t^*+1=2\rho\overset{\text{Obs.~\ref{obs:optimum_cost_after_r3_r4}}}{\geq}\rho\OPT(\sigma_{\leq 3})\,,
	\end{equation*}
	contradicting that \ALG has competitive ratio strictly less than $\rho$.
\end{proof}

We can now prove lower bounds on the competitive ratio of \ALG, depending on which request~$r_5$ is released.

\begin{lemma}
	\label{lem:lower_bound_firstcase}
	If $\pos(2b)<b/2$ and $\servingTime{4}> 2b$, then the adversarial request sequence is well-defined in the sense that $t_{\mathrm m}$ as described in \Cref{def:adversary} exists, and
	\begin{equation*}
		\rho \geq 1+\frac{3b/2}{\rho (t^*+b)-b/2}.
	\end{equation*}
\end{lemma}

\begin{proof}
	Since $\pos(2b)<b/2$ and $r_4=(b,b)$ is not yet served at time $2b$, the server has to cross position $b/2$ at some time $t_{\mathrm{m}}$ after $2b$.
	Therefore, the request~$r_{5}=(t_{\mathrm{m}}, 0)$ is well-defined.

	Between times $b$ and $t_{\mathrm{m}}$, only requests up to $r_4$ are present, so the off\-line optimum cost is~$\OPT(\sigma_{\leq 4})\leq t^*+b$ by \Cref{obs:optimum_cost_after_r3_r4}.
	Since \ALG is $\rho$-competitive, $\ALG[\sigma_{\leq 4}]$ must therefore serve $r_4=(b,b)$ before time~$\rho (t^*+b)$.
	This implies that
	\begin{equation}\label{eq:tm_lower}
		t_{\mathrm{m}}+b/2=t_{\mathrm{m}}+|p_4-\pos(t_{\mathrm m})|\leq \rho (t^*+b)\,.
	\end{equation}
	Since $\pos(t_{\mathrm m})=b/2$ and since $r_4=(b,b)$ and $r_5=(t_{\mathrm{m}}, 0)$ are not yet served by \ALG at time~$t_{\mathrm{m}}$, we have $\ALG(\sigma_{\leq 5})\geq t_{\mathrm{m}}+3b/2$.
	Moreover, $\OPT(\sigma_{\leq 5})=t_{\mathrm{m}}$.
	It follows that
	\begin{equation*}
		\rho \geq \frac{\ALG}{\OPT} \geq \frac{t_{\mathrm{m}}+3b/2}{t_{\mathrm{m}}}
		=1+\frac{3b/2}{t_{\mathrm{m}}}
		\overset{\eqref{eq:tm_lower}}{\geq}
		1+\frac{3b/2}{\rho (t^*+b)-b/2}\,.\qedhere
	\end{equation*}
\end{proof}

To prove a lower bound on the competitive ratio in the second case, we first establish that \ALG must plan to serve $r_3$ before $r_4$, and we derive a lower bound on the time $r_3$ is served.
We prove the claimed serving order by contradiction: assuming the opposite, we can construct an additional request for which the resulting request sequence even yields a lower bound of 1.75.

\begin{lemma}[Serving order of $r_3$ and $r_4$]
	\label{lem:r3_before_r4}
	$\ALG[\sigma_{\leq 4}]$ serves $r_3=(t^*,0)$ before $r_4=(b,b)$.
\end{lemma}

\begin{proof}
	Assume, for the sake of contradiction, that $\ALG[\sigma_{\leq 4}]$ serves the request~$r_4=(b, b)$ before~$r_3=(t^*,0)$.
	We will derive a contradiction to the assumption that $\rho\leq 5/3$ by presenting a request that forces \ALG to have competitive ratio at least $1.75$.

	As a first step, we prove that $\pos(t^*+b)\geq b/2$.
	Suppose for the sake of contradiction that~$\pos(t^*+b)< b/2$.
	At time~$t^*+b<2b$, only requests $r_1$, $r_2$, $r_3$ and $r_4$ have been released, and, by \Cref{obs:optimum_cost_after_r3_r4},~$\OPT(\sigma_{\leq 4})\leq t^*+b$.
	Observe that \ALG must have served $r_4=(b,b)$ before time $t^*+b$ (i.e., $\servingTime{4}\leq t^*+b$), as, otherwise,
	\begin{equation*}
		\ALG(\sigma_{\leq 4})\geq (t^*+b)+(p_4-\pos(t^*+b))+(p_4-p_3)\geq t^*+b+3b/2\,,
	\end{equation*}
	implying that
	\begin{equation*}
		\rho\geq \frac{\ALG(\sigma_{\leq 4})}{\OPT(\sigma_{\leq 4})}\geq \frac{(t^*+b)+3b/2}{t^*+b}= 1+\frac{3b/2}{t^*+b}
		\overset{b> t^*}{>} 1 + \frac{3b/2}{2b} = 1.75\,.
	\end{equation*}
	This contradicts the assumption that $\rho \leq 5/3$.
	Note that $\servingTime{4}\geq t^*+(b-\pos(t^*))$, as the server needs to reach $r_4=(b,b)$ after time $t^*<b$.
	With this, we obtain that
	\begin{align*}
		\pos(t^*+b) & \geq \pos(\servingTime{4})-|(t^*+b)-\servingTime{4}|\\
		             & = b-(t^*+b-\servingTime{4}) \geq b-\pos(t^*).
	\end{align*}
	This yields
	\begin{align*}
		\pos(t^*+b)-\frac{b}{2} & \geq \frac{b}{2}- \pos(t^*) \overset{b> t^*}{>} \frac{t^*}{2}- \pos(t^*)
		= \frac{t^*}{2}- 2\rho +1+t^*
		= \frac{3}{2} t^* -2 \rho +1 \\
		\overset{\text{Lem.~\ref{lem:line_intersection}}} & {\geq} \frac{3}{2} (\rho + (3/(2\rho)) -1) -2\rho +1
		= \frac{9}{4\rho}-\frac{\rho+1}{2}\geq 0,
	\end{align*}
	where the last inequality holds for $\rho\leq (\sqrt{19}-1)/2\approx 1.6794$.
	It follows that~$\pos(t^*+b)\geq b/2$, which completes the first step.

	Define~$k(t)\coloneq(t-t^*)/2$.
	It holds that~$\pos(t^*+b)\geq b/2=k(t^*+b)$.
	Let $s_3$ denote the time when $\ALG[\sigma_{\leq 4}]$ serves $r_3$.
	Then~$\pos(s_3)=p_3=0< k(s_3)$, where~$s_3\geq 2b> t^*+b$, as it takes at least~$2b$ time units to serve $r_4$ before~$r_3$.
	Therefore, there exists a time~$t'\in[ t^*+b, s_3)$ such that~$\pos(t')=k(t')=(t'-t^*)/2$.
	Consider the potential request~$r'=(t', t'-t^*)$.
	\OPT could complete the sequence~$(r_1,r_2,r_3,r_4,r')$ by waiting at~$0$ until time $t^*$, and then moving to $t'-t^*\geq b$ with total cost~$t'$.
	Since \ALG is $\rho$-competitive, it must be able to serve all requests of $(r_1,r_2,r_3,r_4,r')$ before time~$\rho t'$.
	On the other hand, since the server is located at~$\pos(t')=(t'-t^*)/2$ and $s_3>t'$ with $p_3=0$, we have that
	\begin{equation*}
		\ALG((r_1,r_2,r_3,r_4,r'))\geq t'+\frac{3}{2} (t'-t^*)\overset{t'\geq t^*+b\geq 2t^*}{\geq} t'+\frac{3}{2}(t'-\frac{1}{2} t')=1.75t'\,.
	\end{equation*}
	This contradicts the assumption that $\rho\leq 5/3$, which implies that the initial assumption that~$\ALG[\sigma_{\leq 4}]$ serves $r_4=(b,b)$ before $r_3=(t^*,0)$ is false.
\end{proof}

To establish a lower bound on the serving time of $r_3$, we make use of the fact that \ALG must plan to serve $r_2$ before $r_3$ (\Cref{obs:planned_order_r2_r3}).

\begin{lemma}[Serving time of $r_3$]
	\label{lem:serving_time_r3}
	If $b\geq t^*+1/2$, it holds that $\servingTime{3}\geq 3+2t^*-2\rho$.
\end{lemma}

\begin{proof}
	Recall that, by \Cref{obs:planned_order_r2_r3}, $\ALG[\sigma_{\leq 3}]$ serves $r_2=(1,1)$ before $r_3=(t^*, 0)$.
	Moreover, by \Cref{lem:line_intersection}, $t^*<\servingTime{2}$.
	It follows that either $t^*<\servingTime{2}<\servingTime{3}$ or both requests are served after $r_4$ appears, i.e., $t_4\leq \servingTime{2}$ and $t_4\leq\servingTime{3}$.
	If~$t^*<\servingTime{2}<\servingTime{3}$, we obtain, by definition of~$t^*$,
	\begin{equation*}
		\servingTime{3}\geq t^*+|p_2-\pos(t^*)|+|p_3-p_2|\geq t^*+(1-\pos(t^*))+1=3+2t^*-2\rho\,.
	\end{equation*}
	For the case that $t_4\leq \servingTime{2}$ and $t_4\leq\servingTime{3}$, we analyze the position of the server at time~$t_4=b$.
	Since $\ALG[\sigma_{\leq 3}]$ serves $r_2=(1,1)$ before $r_3=(t^*, 0)$, we have that
	\begin{equation*}
		t_4+(1-\pos(t_4))+1
		\leq t_4+|p_2-\pos(t_4)|+|p_2-p_3|
		\leq \ALG(\sigma_{\leq 3})
		< \rho\OPT(\sigma_{\leq 3})
		\overset{\text{Obs.~\ref{obs:optimum_cost_after_r3_r4}}}{\leq} 2\rho\,,
	\end{equation*}
	or, equivalently, $\pos(t_4)> b+2-2\rho$.
	Using $b\geq t^*+1/2$, it follows that
	\begin{equation*}
		\servingTime{3}\geq t_4+|\pos(t_4)-p_3|=b+\pos(t_4) > 2b+2-2\rho \geq 3+2t^*-2\rho\,.\qedhere
	\end{equation*}
\end{proof}

Combining the previous two lemmas, we can prove a lower bound on the competitive ratio for the second case.
Here, the main observation is that it is always optimal for \ALG to serve~$r_3$, then $r_4$ and then $r_5$, which allows us to use the lower bound on $\servingTime{3}$ to obtain a lower bound on the competitive ratio for $\sigma_{\leq 5}$.

\begin{lemma}
	\label{lem:lower_bound_secondcase}
	Let $b\geq t^*+1/2$.
	If $\pos(2b)\geq b/2$ or $\servingTime{4}\leq 2b$, then
	\begin{equation*}
		\rho \geq 1+\frac{3+2t^*-2\rho}{2b}\,.
	\end{equation*}
\end{lemma}
\begin{proof}
	The adversarial sequence concludes by releasing~$r_{5}=(2b, 0)$ at time~$2b$.
	The optimum cost for the entire sequence is $\OPT(\sigma_{\leq 5})= 2b$ (immediately move to $b$ and back to the origin by time $2b$).
	By \Cref{lem:r3_before_r4}, $\ALG[\sigma_{\leq 4}]$ serves $r_3=(t^*,0)$ before $r_4=(b,b)$.
	If~$\servingTime{4}\leq 2b=t_5$, i.e., $r_4$ has already been served when $r_5$ appears, then we have~${\servingTime{3}< \servingTime{4}< \servingTime{5}}$.
	It follows that $\ALG(\sigma_{\leq 5})\geq \servingTime{3}+|p_4-p_3|+|p_5-p_4|=\servingTime{3}+2b$.
	In the other case (i.e., $\servingTime{4}>2b$), we have $\pos(2b)\geq b/2$ and $\servingTime{4}>2b$ by assumption.
	Moreover, we can show that~$\servingTime{3}<2b$, i.e., $r_3$ is served before $r_4$ and $r_5$:
	Otherwise, since~$\ALG[\sigma_{\leq 4}]$ serves $r_3$ before $r_4$ (\Cref{lem:r3_before_r4}), it holds that
	\begin{equation*}
		2\rho b \overset{b\geq t^*}{\geq} \rho(t^*+b)
		\overset{\text{Obs.~\ref{obs:optimum_cost_after_r3_r4}}}{\geq} \rho\OPT(\sigma_{\leq 4})
		\geq\ALG(\sigma_{\leq 4})
		\geq 2b+|\pos(2b)-p_3|+|p_4-p_3|\geq 7b/2\,,
	\end{equation*}
	contradicting $\rho\leq 5/3$.
	Since $\pos(2b)\geq b/2$ and $\servingTime{4}>2b$, the fastest way for \ALG to complete the sequence~$\sigma_{\leq 5}$ is to serve $r_4=(b,b)$ before $r_{5}=(2b,0)$.
	As before, we obtain~$\ALG(\sigma_{\leq 5}) \geq \servingTime{3}+2b$.
	It follows that
	\begin{equation*}
		\rho \geq \frac{\ALG(\sigma_{\leq 5})}{\OPT(\sigma_{\leq 5})}\geq \frac{\servingTime{3}+2b}{2b}
		=1+\frac{\servingTime{3}}{2b} \overset{\text{Lem.~\ref{lem:serving_time_r3}}}{\geq} 1+\frac{3+2t^*-2\rho}{2b}\,.\qedhere
	\end{equation*}
\end{proof}

To prove \Cref{thm:lower_bound}, we combine the bounds of \Cref{lem:lower_bound_firstcase} and \Cref{lem:lower_bound_secondcase} and choose $b$ such that both bounds are equal.
To ensure that the lower bounds still apply, we first verify that this choice of $b$ is valid in the sense that $b\geq t^*+1/2$.

\begin{lemma}
	\label{lem:b_geq_t_plus_half}
	Let $\rho\in[3/2,5/3]$ and $t^*\geq \frac{7}{5}$.
	If $b$ satisfies
	\begin{equation}
		\label{eq:b_balancing}
		\frac{3+2t^*-2\rho}{2b}=\frac{3b/2}{\rho (t^*+b) - b/2}
	\end{equation}
	then $b\geq t^*+1/2$.
\end{lemma}

\begin{proof}
	Notice that the left side of \eqref{eq:b_balancing} is decreasing in~$b$ and the right side of \eqref{eq:b_balancing} is increasing in~$b$, as
	\begin{equation*}
		\frac{3b/2}{\rho(t^*+b)-b/2}=\frac{3}{2\rho-1+\frac{2\rho t^*}{b}}
	\end{equation*}
	and $\rho t^*>0$.
	For $b=t^*+\frac{1}{2}$, the left-hand side becomes
	\begin{align*}
		L\coloneq \frac{3+2t^*-2\rho}{2b} & =\frac{3+2t^*-2\rho}{2t^*+1}=1+\frac{2-2\rho}{2t^*+1}
	\end{align*}
	and the right-hand side becomes
	\begin{align*}
		R\coloneq \frac{3b}{2\rho(t^*+b)-b}
		&=\frac{3(t^*+1/2)}{(4\rho-1)(t^*+1/2)-\rho}
		=\frac{3}{4\rho-1-\frac{\rho}{t^*+1/2}}\,.
	\end{align*}
	We want to show that $L > R$, as this implies that both sides can become equal only for some~$b > t^*+\frac{1}{2}$.
	Notice that, as $2-2\rho < 0$ and $\rho>0$, $L$ is increasing in $t^*$ and $R$ is decreasing in~$t^*$.
	Therefore, it suffices to show $L > R$ for $t^*=7/5$, which is the smallest possible value of~$t^*$.
	For this choice of $t^*$, we have
	\begin{equation*}
		L=\frac{29-10\rho}{19}\qquad R=\frac{57}{66\rho-19}\,,
	\end{equation*}
	so $L>R$ reduces to
	\begin{equation*}
		0<(29-10\rho)(66\rho-19) -57\cdot 19 =-660\rho^2+2104\rho-1634
	\end{equation*}
	where the right-hand side has roots $\rho\approx 1.339$ and $\rho\approx 1.849$ and negative leading coefficient.
	Therefore, it is positive for $\rho\in[1.34, 1.84]\supset [3/2, 5/3]$.
\end{proof}

The resulting lower bound on the competitive ratio depends on the value $t^*$ chosen by the algorithm.
To find an unconditional lower bound, we prove that it is increasing in $t^*$, and determine its value for the smallest possible $t^*$.
To this end, we use the following intermediate lemma, stating that our choice of $b$ to make the bounds of \Cref{lem:lower_bound_firstcase} and \Cref{lem:lower_bound_secondcase} equal is increasing in $t^*$.

\begin{lemma}
	\label{lem:b_increasing_in_t}
	Assume that $\rho\in(3/2,5/3]$ is fixed and $b(t)$ is maximal such that
	\begin{equation}
		\label{eq:b_balancing_t}
		\frac{3+2t-2\rho}{2b(t)}=\frac{3b(t)}{2\rho(t+b(t)) - b(t)}\,.
	\end{equation}
	Then $b(t)$ is increasing and differentiable in $t$ for all $t\geq 7/5$.
\end{lemma}

\begin{proof}
	The function $b(t)$ can be determined by reformulating \eqref{eq:b_balancing_t}, where we note that the denominators are non-zero, as $b(t)\geq t+1/2> 0$ by \Cref{lem:b_geq_t_plus_half}:
	\begin{align*}
		     & \frac{3+2t-2\rho}{2b(t)}=\frac{3b(t)}{2\rho(t+b(t)) - b(t)}                          \\
		\Leftrightarrow~& (3+2t-2\rho)(2\rho(t+b(t)) - b(t))=6b(t)^2                                           \\
		\Leftrightarrow~& (-4\rho^2 + 4\rho t + 8\rho - 2t-3)b(t)-4\rho^2 t + 4\rho t^2 + 6\rho t = 6b(t)^2\,,
	\end{align*}
	Note that $3+2t-2\rho>0$, as $t\geq 7/5$ and $\rho\leq 5/3$.
	Hence, the two roots of this quadratic equation in $b(t)$ have product $-2\rho t(3+2t-2\rho)/6<0$, so $b(t)$ is a simple root and therefore differentiable in $t$.

	To show that $b(t)$ is increasing, we consider the two sides of \eqref{eq:b_balancing_t} as functions of $t$ and $b$, i.e.,
	\begin{equation*}
		L(t,b)\coloneq\frac{3+2t-2\rho}{2b}\quad\text{and}\quad R(t,b)\coloneq\frac{3b}{2\rho(t+b)-b}=\frac{3}{2\rho-1+\frac{2\rho t}{b}}\,.
	\end{equation*}
	For fixed $b>0$, $L$ is increasing in $t$ and $R$ is decreasing in $t$.
	Hence, for $t'>t$, we have
	\begin{equation*}
		L(t',b(t))>L(t,b(t))=R(t,b(t))>R(t',b(t))\,.
	\end{equation*}
	Moreover, as in the proof of \Cref{lem:b_geq_t_plus_half}, for fixed $t'$, $L$ is decreasing in $b$ and $R$ is increasing in~$b$.
	Thus, $L(t',b)-R(t',b)$ is decreasing in $b$, positive for $b=b(t)$ and zero for $b=b(t')$, which implies $b(t')>b(t)$.
\end{proof}

Using this lemma, we can now show that the lower bound itself increases in $t^*$.

\begin{lemma}
	\label{lem:lb_increasing_in_t}
	Assume that $\rho\in(3/2,5/3]$ is fixed and $b(t)$ is maximal such that
	\begin{equation*}
		\frac{3+2t-2\rho}{2b(t)}=\frac{3b(t)}{2\rho(t+b(t)) - b(t)}\eqcolon \LB(t).
	\end{equation*}
	Then $\LB(t)$ is increasing in $t$ for $t\geq 7/5$.
\end{lemma}

\begin{proof}
	Computing the derivative of $\LB(t)$ using the left side yields
	\begin{align*}
		\frac{\mathrm{d}}{\mathrm{d}t}\LB(t) & =\frac{4b(t) - 2b'(t)(3+2t-2\rho)}{(2b(t))^2}=\frac{4(b(t)-tb'(t))+2b'(t)(2\rho-3)}{(2b(t))^2}\,,
	\end{align*}
	while the right side yields
	\begin{align*}
		\LB'(t) & =\frac{3b'(t)(2\rho(t+b(t))-b(t))-3b(t)(2\rho(1+b'(t)) - b'(t)) }{(2\rho(t+b(t))-b(t))^2}                             \\
		        & =\frac{6b'(t)\rho t + 6b'(t)\rho b(t)-3b'(t)b(t) - 6\rho b(t) - 6\rho b'(t)b(t) + 3b'(t)b(t)}{(2\rho(t+b(t))-b(t))^2} \\
		        & =\frac{6\rho(tb'(t) - b(t))}{(2\rho(t+b(t))-b(t))^2}\,.
	\end{align*}
	It follows that
	\begin{equation*}
		\sgn(\LB'(t)) = \sgn(tb'(t) - b(t)) = \sgn\left(2(b(t)-tb'(t)) + (2\rho - 3)b'(t)\right).
	\end{equation*}
	If $tb'(t)-b(t)> 0$, then $\LB'(t)>0$ by the first equality.
	Otherwise, we have~$b(t)-tb'(t)\geq 0$.
	Using~$\rho> 3/2$ and~$b'(t)\geq 0$ (\Cref{lem:b_increasing_in_t}), we obtain that~$(2\rho - 3)b'(t)\geq 0$, hence $\LB'(t)\geq 0$ by the second equality.
\end{proof}

Combining the previous lemmas, we can now prove \Cref{thm:lower_bound}.

\begin{proof}[Proof of \Cref{thm:lower_bound}]
	\Cref{lem:lower_bound_firstcase,lem:lower_bound_secondcase} combined yield that, for every~$\rho\in(3/2,5/3]$ and every~$b\geq t^*+1/2$, we have
	\begin{equation}
		\label{eq:lower_bounds_summary}
		\rho \geq 1+ \min \left(
		\frac{3b/2}{\rho (t^*+b)-b/2}
		,
		\frac{3+2t^*-2\rho}{2b}
		\right)\eqcolon \LB(t^*,\rho, b).
	\end{equation}

	The remainder of the proof is purely analytical and dedicated to showing that, for every~${\rho\in (3/2,1.6463]}$ and $t^*\geq \rho + 3/(2\rho) -1$, there exists $b\geq t^*+1/2$ such that $\LB(t^*,\rho,b)>\rho$, contradicting \eqref{eq:lower_bounds_summary}.

	To this end, we choose the maximal $b(t^*, \rho)$ such that the two values in the minimum coincide.
	By \Cref{lem:line_intersection}, we have that
	$t^*\geq \rho+\frac{3}{2\rho}-1\geq \frac{3}{2} + \frac{3}{2} \cdot \frac{3}{5}-1 = 7/5$, and therefore, \Cref{lem:b_geq_t_plus_half} gives~$b(t^*, \rho)\geq t^*+1/2$, so that the lower bound still applies.
	Moreover, it follows from \Cref{lem:lb_increasing_in_t} that the resulting lower bound $\LB(t^*, \rho, b(t^*, \rho))$ is increasing in $t^*$.
	Therefore, we obtain a general lower bound by setting $t^*=\rho+3/(2\rho) -1$ (since we have proven in \Cref{lem:line_intersection} that $t^*$ is lower bounded by this value).
	Plugging this in, we obtain
	\begin{equation*}
		\rho \geq 1+ \min \left( \frac{3b}{2\rho^2-2\rho + 2\rho b + 3-b}, \frac{3+\rho}{2\rho b} \right)
	\end{equation*}
	resulting in
	\begin{equation*}
		b=\frac{2\rho^2 + 5\rho - 3 +\sqrt{52\rho^4 + 116\rho^3 - 59\rho^2 +186\rho + 9}}{12\rho}
	\end{equation*}
	and a lower bound on the competitive ratio of
	\begin{equation*}
		1+\frac{6(\rho+3)}{2\rho^2+5\rho-3 + \sqrt{52\rho^4 + 116\rho^3 - 59\rho^2 +186\rho + 9}}\,,
	\end{equation*}
	which decreases in $\rho>0$ and equals $\rho$ for some $\rho> 1.6463$.
	Hence, for every $\rho\in(3/2,1.6463]$, this lower bound is strictly larger than~$\rho$, a contradiction.
	Therefore, no algorithm has competitive ratio less than~$1.6463$.
\end{proof}

\section*{Disclosure of AI usage}
The results presented in this paper and its preliminary version \cite{Baligacs2026} were obtained without any AI assistance. 
In preparing the present paper, the authors used Claude Opus 5.5 for minor language edits.

\bibliographystyle{plainurl}
\bibliography{HalfLineTSP}

\end{document}